# Low-temperature surface conduction in the Kondo insulator SmB$_6$

Steven Wolgast,[1] Çağlıyan Kurdak,[1] Kai Sun,[1] J. W. Allen,[1]
Dae-Jeong Kim,[2] and Zachary Fisk[2]

[1] *Department of Physics, Randall Laboratory, University of Michigan,
Ann Arbor, Michigan 48109, USA*
[2] *Department of Physics and Astronomy, University of California at Irvine,
Irvine, California 92697, USA*

We study the transport properties of the Kondo insulator SmB$_6$ with a specialized configuration designed to distinguish bulk-dominated conduction from surface-dominated conduction. We find that as the material is cooled below 4 K, it exhibits a crossover from bulk to surface conduction with a fully insulating bulk. We take the robustness and magnitude of the surface conductivity, as is manifest in the literature of SmB$_6$, to be strong evidence for the topological insulator metallic surface states recently predicted for this material.

Kondo insulators [1–6] are mixed valent or heavy fermion f-electron compounds that manifest not only the quenching of *f*-electron magnetic moments but also a low-temperature resistivity rise that implies a gap at the Fermi energy $E_F$. Both phenomena result from hybridization between conduction electrons and the strongly interacting *f*-electrons. Application [1,2] of Fermi liquid concepts leads to renormalized heavy quasi-particles that can be fully gapped for appropriate band symmetries. SmB$_6$ is a classic example that helped to start the field of rare earth mixed valence [7–9]. With decreasing temperature (*T*) its resistivity and the magnitude of its Hall effect increase exponentially as expected for thermally activated transport across a gap [10,11]. Optical [12,13], photoemission [14,15] and tunneling [16] spectroscopy do indicate gap or pseudogap features, in particular small gaps consistent with the measured transport activation energy (2.5-3.5 meV) [10,11]. However at lower temperatures (*T* < 4 K) the resistivity and Hall effect do not continue to diverge but saturate [10,11] and remain finite as $T$ goes toward 0 K, implying that $E_F$ lies in conducting states within the small gap. Assuming that the transport is dominated by the three-dimensional (3D) bulk, these in-gap states and the residual resistivity remain a mystery after more than 30 years. On the one hand, the resistivity systematically increases as samples are made more stoichiometric [17], which suggested [10] an impurity band, but in the best samples the resistivity exceeds the Mott impurity band limit [18] by at least a factor of 15 [10]. On the other hand, the pressure dependence of the transport suggested [11] intrinsic in-gap states, but the resistivity then exceeds the limiting value corresponding to a scattering center in every unit cell, scattering at the unitarity limit [19,20], by at least a factor of 80 [11].

The solution to this mystery may lie in a recent theoretical prediction [21–24] that Kondo insulators can also be topological insulators. Topological insulators, which have been the subject of intense theoretical investigation over the last several years [25–34], are insulating in the bulk but have in their gaps metallic surface states (edge states in two-dimensional (2D) topological insulators) in the form of Dirac cones with helical spin structures. These states are robust, being protected by time-reversal symmetry characterized by (strong or weak) $Z_2$ topological indices. The theory was initially developed for weakly correlated band insulators, but within the framework of a Fermi liquid description [1,2], Kondo insulators can have the same symmetry-protected topological properties [21–24], and $SmB_6$ is predicted [21,24,35,36] to be a strong topological insulator. If topologically protected metallic surface states give rise to the residual low-temperature resistivity, all the difficulties with fundamental resistivity limits described above for 3D bulk explanations would be evaded.

3D topological insulators have been confirmed and studied experimentally in such materials as $Bi_{1-x}Sb_x$ [37], $Bi_2Se_3$ [38,39], and $Bi_2Te_3$ [39]; however, transport characterization has been very challenging because the bulk of these materials is conductive. Various strategies to suppress bulk conductivity, e.g., studying thin films, gating, and doping, have been employed, and sophisticated theoretical arguments are used to infer the success of these strategies. In strained HgTe [30,40], the quantum Hall effect has been reported for thin films with a thickness of 70 nm, where the surface contribution clearly dominates the Hall signal at millikelvin temperatures [41]. Quantum oscillations have also been observed in 3D bulk samples, e.g., $Bi_2Te_2Se$ and doped $Bi_2Te_2Se$, which have a large bulk resistivity. However, the surface may contribute no more than 70% of the total conductance in these samples [42,43]. In general, identifying a material with a topologically protected surface and a fully insulating bulk would greatly simplify the study of the surface states for many important bulk-sensitive techniques. We present here strong evidence and arguments that $SmB_6$ is that material.

To solve the mystery of the residual resistivity and to test the prediction of the presence of topologically protected conducting surface states in this material, we performed transport experiments with a specialized sample geometry shown in Fig. 1(a) [44]. Details on the fabrication of the real sample (Fig. 1(b)) may be found in the Supplemental Material [45]. We use a thin sample of $SmB_6$ with eight coplanar electrical contacts on the (100) and (-100) surfaces, four on each side, to determine whether the conduction is dominated by bulk or surface. The top and bottom leads were aligned along the [001] direction. If the material is an isotropic bulk conductor [45], the four-terminal resistance of the sample would be proportional to the bulk resistivity of the material with a proportionality constant that depends only on the geometry of the sample bulk and the contacts. However, if there is a crossover from the bulk to the surface, the relative contributions from the bulk and surface resistivities can be suppressed or exaggerated, depending on the position of the current and voltage leads.

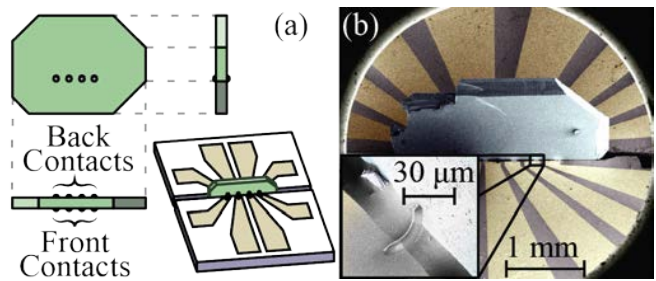

**FIG 1.** (Color online) **(a)** A schematic diagram of a piece of $SmB_6$ with eight coplanar contacts, four on each side, sandwiched between two silicon wafer pieces with gold contact pads. **(b)** A scanning electron microscope image of a single crystal of $SmB_6$, sandwiched between two silicon wafer pieces with lithographically defined contact pads. **Inset**: a close-up of one of the platinum contacts connecting the $SmB_6$ to a gold contact pad.

When we perform conventional four-terminal resistance measurements, just using the contacts on the front surface of the sample, we obtain data shown in Fig. 2, which is consistent with previous measurements of $SmB_6$ [10,11], featuring an insulator-like increase in resistance with decreasing temperature, but with a weakly temperature-dependent plateau at low temperatures. We can model the measured conductance as having two independent contributions: $G_{measured} = G_{insulator} + G_{plateau}$ with $G_{insulator} = G_a \exp(-\Delta/k_B T)$. We then extract $G_{insulator}$ down to 3 K. A linear fit of the Arrhenius plot gives us an activation energy $\Delta$ of 3.47 meV, which is consistent with previously published measurements of $SmB_6$ [10,11].

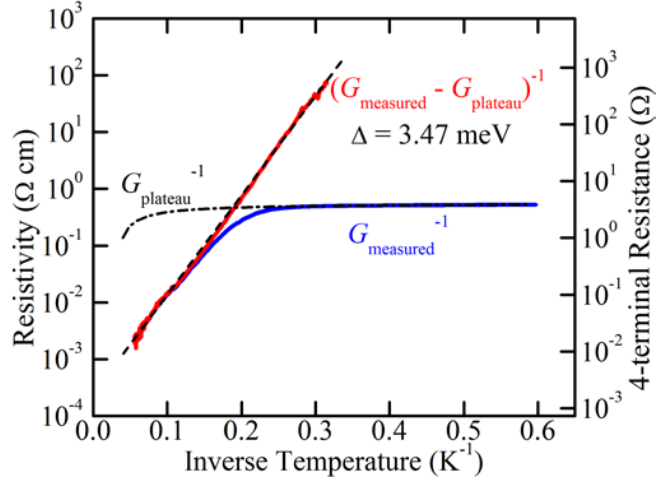

**FIG 2.** (Color online) A logarithmic plot of 3D resistivity $G_{measured}$ (solid blue line) vs inverse temperature. A linear model of the plateau conductance $G_{plateau}$ (dash-dotted line) is removed from $G_{measured}$ to extrapolate (red line) the bulk resistivity to 3 K. A linear fit (dashed line) yields an activation energy of 3.47 meV.

The conventional lateral measurement $R_{lat}$ using contacts from one side cannot distinguish whether the conduction at low temperatures is bulk-dominated or surface-dominated. However, we can explore other measurement configurations using contacts from both sides; specifically, we can make a vertical measurement $R_{vert}$ by passing current from one front-side contact to the back-side contact directly opposite, and measuring the voltage using a different set of opposing front-side and back-side contacts. We can also make a hybrid measurement $R_{hyb}$ by passing current through two front-side contacts as in the lateral measurement, but measuring the voltage on two back-side contacts. Cross sections of these configurations, derived from finite element analysis simulations, are illustrated in Fig. 3. If the plateau is a bulk transport phenomenon, the resistance will be proportional to the resistivity for all three measurement configurations, each with a different proportionality constant. In other words, the temperature dependencies of $R_{lat}$, $R_{vert}$, and $R_{hyb}$ normalized to their respective room temperature values are expected to be identical. However, if the plateau is due to surface conduction, these three four-terminal resistances behave dramatically differently as a function of temperature.

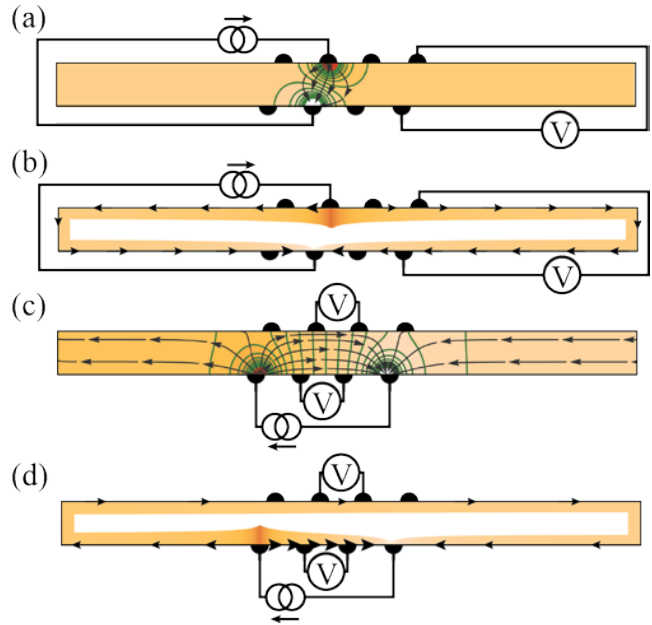

**FIG 3.** (Color online) A cross section of the sample along the electrical contacts. Arrows indicate current direction; lines indicate equipotentials. **(a)** Current passes vertically through the bulk, far away from the voltage contacts. **(b)** The bulk in (a) becomes insulating, forcing the current to flow around the edge. The surface potential is indicated by the thickness of the orange region. **(c)** Current passes laterally through the bulk, and the front-side and back-side voltages are measured at similar equipotentials. **(d)** The bulk in (c) becomes insulating, isolating the back-side contacts from the majority current flow.

In the vertical configuration at high temperature, nearly all the current will flow vertically directly through the sample if the bulk is conductive, as shown in Fig. 3(a). Because the voltage contacts are located far away from the current, there is virtually no current near the voltage contacts, and $R_{vert}$ is unmeasurably small. For this reason, such a configuration is never used to measure an ordinary sample. Even though the resistivity increases significantly at low temperatures, the current will continue to flow in this configuration as long as the bulk is conductive. However, if the material becomes a surface-conductor at low temperatures, the entire current will be forced to flow around the long dimensions of the sample (Fig. 3(b)). In this case, the voltage contacts are very close to the current contacts, compared to the total current path around the edges; thus, $R_{vert}$ will become very large. Meanwhile, in the lateral configuration shown in Fig. 3(c), $R_{hyb}$ should be nearly identical to $R_{lat}$ at high temperatures when the bulk is conducting. This is because the current is nearly uniform between the front-side and back-side contacts. Again, if the bulk remains conducting at low temperatures where the resistivity becomes large, the current will still follow the same path. However, if the material becomes a surface conductor (Fig. 3(d)), the back-side contacts become electrically remote from all the front-side contacts. Most of the current will flow only along the front side, and very little current will take the long path around the back side; thus $R_{hyb}$ will be much smaller than $R_{lat}$.

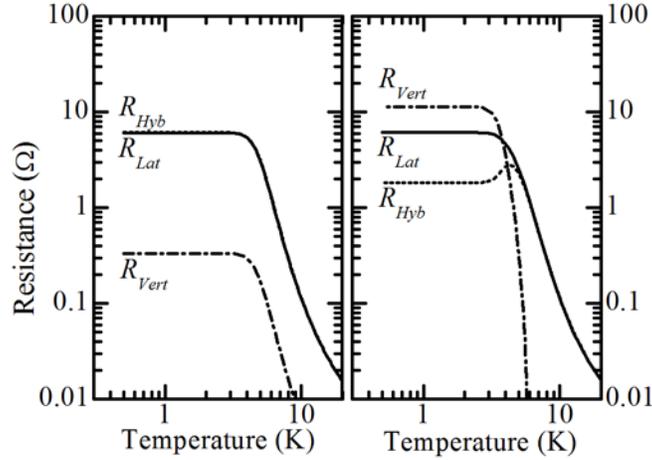

**FIG 4.** Simulated log-log plots of the four-terminal resistances as a function of temperature. **Left**: the case where the saturation conductivity is a bulk phenomenon. **Right**: the case where the saturation conductivity is a surface conductivity.

The finite element analysis simulated the electric potential as a function of temperature for both the bulk-only scenario and the bulk-surface crossover scenario. We assumed a system with an isotropic conductivity $\sigma_{insulator} = \sigma_a \exp(-\Delta/k_B T)$ in parallel with a constant conductivity $\sigma_p$, with $\sigma_a$ and $\sigma_p$ tuned to provide a resistance that is qualitatively similar to our lateral measurements. In the bulk-only scenario, plotted in Fig. 4(a), we assume that these competing conductivities are both bulk phenomena. Because the current flow pattern does not change in this scenario, the bare resistances scale uniformly, each proportional to the resistivity, but with different proportionality constants. In the bulk-surface crossover scenario, we associate $\sigma_{insulator}$ with the bulk, and $\sigma_p$ becomes a sheet conductivity associated with the surface (Fig. 4(b)). Here, the measurements scale uniformly far above and below a crossover temperature, but near the crossover temperature, where $\sigma_p \approx \sigma_{insulator} \times t$ ($t$ is the thickness of the sample), the measurements do not scale with each other at all, and the proportionality relation is broken. We notice that $R_{lat}$ is qualitatively very similar in both configurations, making it difficult to distinguish between bulk-dominated and surface-dominated conduction from this measurement alone. We also note that in the surface-conductor case, $R_{hyb}$ exhibits a clear peak near the crossover temperature before settling to a smaller low-temperature value, as predicted. Finally, the change in $R_{vert}$ near the crossover temperature is dramatically faster than the changing resistivity.

    To measure the real sample, we performed standard lock-in measurements in the configurations described above. $R_{lat}$, $R_{hyb}$, and $R_{vert}$ are plotted in Fig. 5. The measurements behave remarkably like the crossover case of the simulations, with a distinct peak in $R_{Hyb}$ at 3.8 K, demonstrating conclusively that $SmB_6$ becomes a surface conductor below this temperature. We note, in particular, that $R_{lat}$ and $R_{hyb}$ scale with each other on each side of the crossover regime, suggesting the current path remains fixed, but they diverge near the crossover temperature, indicating a change in the current path. We also note that $R_{vert}$ increases dramatically as the temperature drops below the crossover, even more than predicted in the simulation. We attribute this discrepancy to geometrical differences between the simulated slab surface and the real sample surface. These features cannot be explained by bulk conduction in any cubic system, even with an anisotropic conductivity.

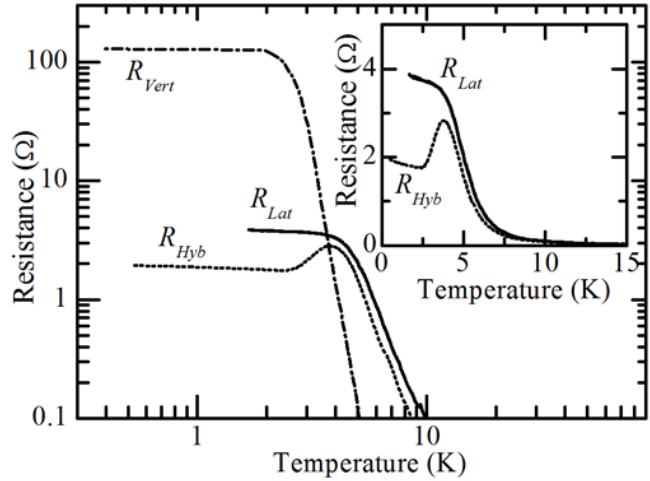

**FIG 5.** A log-log plot of $R_{lat}$ (solid), $R_{vert}$ (dash-dotted), and $R_{hyb}$ (dotted) as a function of temperature. **Inset**: a linear plot of $R_{lat}$ and $R_{hyb}$, emphasizing the divergence between them between 3 and 5 K.

Our experiments prove unambiguously that as temperature is reduced, the system turns from a 3D bulk conductor into a 2D surface conductor with an insulating bulk. They resolve the long-standing puzzles surrounding $SmB_6$ at low temperatures, which are caused by treating the low-temperature conductivity as a bulk conductivity. In fact, the "resistance ratio," which has conventionally been used to assess the quality of $SmB_6$ crystals, is now expected to be non-universal, depending on the bulk stoichiometry, the surface quality, and the sample thickness, which are all independent parameters.

We measured a van der Pauw-like sample of $SmB_6$ and obtained a remarkably low sheet resistance of less than 10 Ω, which is consistent with prior measurements in the literature [10,11] and competitive with the most conductive two-dimensional electron systems and thin films ever developed, as we present in the Supplemental Material [45]. Based on our data, we must now look at many of the previous transport experiments performed on this material system from the perspective of pure surface transport. In particular, the proposed transport mechanism must explain the remarkably low resistivity of the surface, the large carrier density extracted from Hall measurements, and the many-orders-of-magnitude change of the saturation Hall resistivity with pressure up to 45 kbar [11].

We now assess the likelihood of some non-topological origin of the surface conductivity. We note first that previous surface studies of bulk samples came to highly ambiguous conclusions because "surface" was viewed implicitly as "extrinsic and fragile." Deliberate oxidation after polishing [47] modestly decreased but did not eliminate the residual conductivity, leading to the conclusions that the "saturation behavior is a bulk property" but also that "surface metallic conduction plays a partial role." Based on a finding that etching modestly increased the residual resistivity, Kebede *et al.* [48] proposed that it arises from a residual conducting layer ("surface crud"). Significantly however, even after etching away 10%-30% of the samples' weights, the residual conductivity remained and the authors concluded that "the surface crud is not a discrete layer but rather is continuously changing spatially. "Another study [49] of the effect of polishing an etched surface concluded that the residual conductivity has a "dominant bulk nature" but that surface states "make a significant contribution." These studies clearly show that the residual conductivity is robust against the surface treatments employed. Our experiments show that the residual conductivity definitely arises from the surface, and our topological insulator (TI) explanation, in which the conduction is nonetheless intrinsic, is capable of rationalizing the past experience. [50]

One might consider electrical transport through native surface instabilities such as surface reconstruction [51,52], a non-TI metallic surface state [53], a band inversion layer [54], or a change of valence relative to the bulk [49]. It is well known that such surface phenomena are highly vulnerable to surface contamination and are only observed on surfaces cleaned and maintained in ultrahigh vacuum (UHV). Given the long $SmB_6$ history of a robust conductivity measured on samples in air and with contacts prepared in a variety of ways, along with the remarkably high conductivity of our samples, the hypothesis of conduction due to such unrobust phenomena is far less *a priori* credible than the hypothesis of conduction due to surface states protected by very general topological properties that are not in question for the electronic structure of a gap in this material. Furthermore, if the low *T* saturation Hall coefficient is interpreted in the context of an ordinary single-band surface conduction, the resulting carrier density is unphysically large for any 2D system, and its pressure dependence [11] implies orders-of-magnitude changes in thickness for any 3D system, which is also unphysical. This rules out the possibility of a band inversion layer or a residual metal surface layer such as aluminum coming from the flux growth method. (See the Supplemental Material for a more detailed explanation. [45])

We cannot of course claim to have excluded literally all other conceivable possibilities for the surface conduction, but having recognized the fragility of normal surface states as a major barrier, and having ruled out a residual conducting layer or a conventional surface two-dimensional electron gas, there is a very strong motivation to turn to the TI scenario. We have already noted that the literature experience with surface treatments fits elegantly into the fundamental rubric that TI surface states have a protected status not enjoyed by ordinary surface states. The suppression of backwards scattering arising from the helical spin structure of topological insulators may help to explain the remarkably high conductivity measured in $SmB_6$. A recent calculation [55] suggests that there are multiple Dirac cones in the energy gap of the (100) surface. In this case, a multipocket transport model must be used, and the Hall coefficient alone is insufficient to determine the carrier density of the material. Ambipolar conduction has cancellation effects which diminish or even reverse the sign of the Hall coefficient. The observed pressure dependence of the Hall coefficient would then imply a change in the balance between electrons and holes as the gap is steadily reduced by pressure. TI theory thus provides a way to understand the pressure dependence of the transport measurements, but additional measurements of quantum oscillations or with gating techniques will be required to determine the carrier density.

We conclude that the best current working hypothesis is that $SmB_6$ is indeed a topological insulator [56]. We note that such TI surface states in $SmB_6$ can be easily studied on bulk samples with no special effort to suppress bulk conductivity beyond cooling below a readily achievable 3.8 K. Understanding the basic transport properties of this system will require a great effort over a very broad range of experiments, but may allow many existing theories [57] to be tested. Furthermore, this material, a strongly correlated 3D topological state of matter, opens new opportunities to study the interplay between strong-correlation effects and topology in the search for new quantum phases, new quantum phase transitions, and new principles of physics.

ACKNOWLEDGMENTS

The authors wish to acknowledge Richard Field III for his photography services, Yun Suk Eo for assisting with the experiment, Ilya Vugmeyster for help with polishing, Jason Cooley for sample preparation advice, Emanuel Gull for helpful discussions on numerical techniques, and Piers Coleman and Shou-Cheng Zhang for advice on improving the manuscript. This project was performed in part in the Electron Microbeam Analysis Laboratory (EMAL), which is supported by NSF Grant No. DMR-0320740, and in the Lurie Nanofabrication Facility (LNF), a member of the National Nanotechnology Infrastructure Network, which is supported by the National Science Foundation. The portion performed at the University of Michigan was supported by NSF Grant No. DMR-1006500. The portion performed at the University of California at Irvine was supported by NSF Grant No. DMR-0801253.

**Supplemental Information**

## SAMPLE PREPARATION

For our experiment, we selected a SmB$_6$ crystal measuring 2 × 1.5 mm and polished it to a thickness of 160 μm, finishing with P2400 grit paper. The sample was placed into nitric and hydrochloric acids to remove aluminum flux remaining on the surface from crystal growth. For contacts, we used two 500-μm silicon wafer pieces with a 300-nm silicon oxide insulating later and lithographically patterned gold contact leads with 250-μm spacing. The pieces were cleaved across the leads, providing a flat edge on each piece to which the gold leads extended. The SmB$_6$ sample was sandwiched between the cleaved edge of the wafer pieces and glued into place with Varian Torr Seal. The gap between the sample and the cleaved surface edge varied from < 5 to 30 μm. We observed epoxy wetting in portions of the gaps.

We deposited platinum contact wires connecting the SmB$_6$ to the gold leads using ion beam-induced deposition. The ion beam was incident on the sample at 52°. The wires were 3 μm wide and 1–3 μm thick, with a length sufficient to span the gap between the SmB$_6$ and the gold lead, varying from 20 to 50 μm. The epoxy wetting served as an insulating "bridge" for the platinum between the SmB$_6$ and some of the gold leads. It was possible to deposit a wire in a few places without the epoxy bridge, provided the gap was smaller than 10 μm; a few of these were reliable and Ohmic even at cryogenic temperatures.

## HALL MEASUREMENTS

We constructed a sample with a van der Pauw-like geometry. The sample is 3.47 mm × 1.32 mm × 170 μm, and the two large faces were polished with P4000 grit paper.

The sample was mounted to a glass substrate with Varian Torr Seal, and indium contacts were placed along the edge (Fig. S1). Two leads extended along the short edges to function as current leads, while four additional leads were placed along the long edges, two on each side, for voltage contacts.

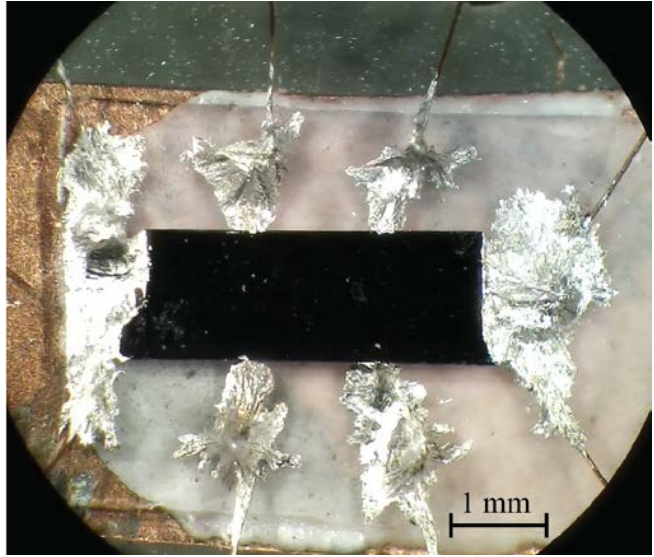

**Figure S1** – Image showing the contact positions on the van der Pauw sample.

This configuration allows us to easily determine the conductivity of the sample with two voltage leads on the same edge using lock-in techniques at 26.6 Hz in a $^3$He cryostat. We obtained a remarkably low sheet resistance of 9.1 Ω below the crossover temperature. This configuration also allows us to measure the Hall conductivity by using opposing leads on opposite edges and sweeping the magnetic field perpendicular to the flat surfaces of the sample. Our Hall measurements were abnormal, with a barely detectable slope and a large temperature-dependent feature near 0-field (Fig. S2). As the temperature is lowered, the Hall coefficient rises in a manner consistent with insulators, but it suddenly drops about two orders of magnitude at the crossover temperature and plateaus at lower temperatures (Fig. S2(b)). This is difficult to interpret in the context of bulk conduction, but is completely expected in the surface conduction scenario. This behavior is qualitatively consistent with prior measurements of SmB$_6$ [S1, S2], though the quantitative details of the crossover and the plateau are expected to depend heavily upon the surface quality and preparation. For our sample, the Hall coefficient above the crossover temperature is about a factor of 10 smaller than that reported by Cooley *et al.* [S2], and at the lowest temperatures is less than $4\times10^{-4}$ Ω/T. The small-field peak is strong in our Hall measurements, but 10 times weaker in our magnetoresistance measurements. In fact, measuring the magnetoresistance using the contacts on the opposite edge of the sample changes the direction of the peak. This suggests that the feature may be associated with the Hall measurement rather than the magnetoresistance measurement, or that the surface conduction may be non-uniform and uncontrolled.

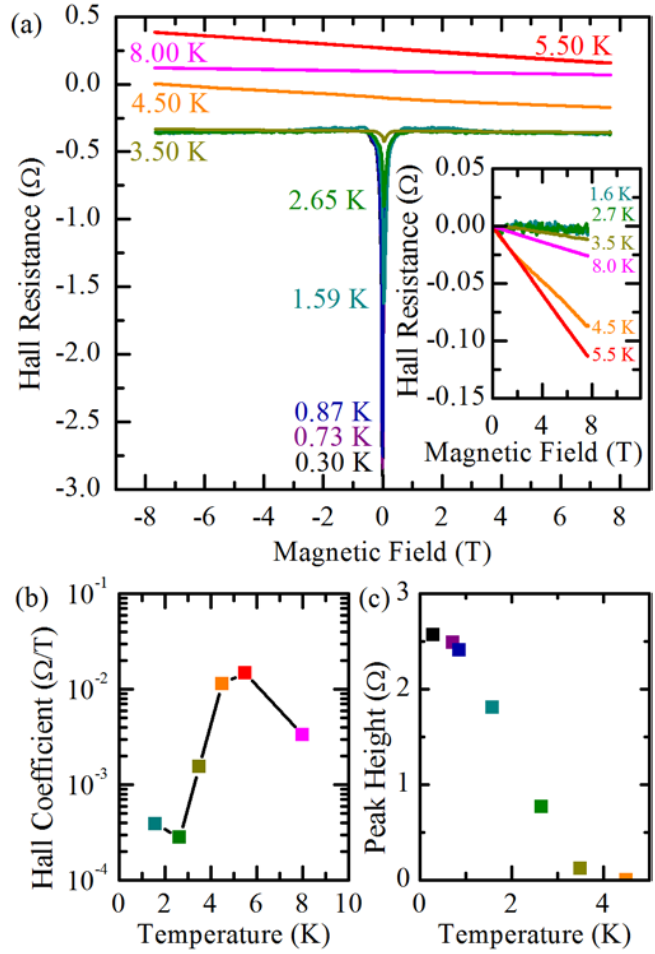

**Figure S2 – (a)** Plots of bare Hall resistance at several temperatures as a function of magnetic field. At the crossover, the Hall slope disappears, and a 0-field feature appears. **Inset** Antisymmetrized Hall resistance at several temperatures as a function of magnetic field. **(b)** The Hall coefficient as a function of temperature. **(c)** A plot of the feature peak height as a function of temperature.

ORDINARY SURFACE STATE MECHANISMS

The most conductive ordinary 2D surface states are due to a band inversion mechanism. In some low-bandgap semiconductors such as InAs, $E_F$ is pinned at the conduction band, leading to localized charges at the surface and a significant bending of the conduction band. A two-dimensional electron layer (inversion layer) can be formed a few nanometers below the surface of the semiconductor in a triangular confinement potential generated by the electrical field due to localized surface states. The researchers working on field effect transistors have tried to attain high carrier density two-dimensional electron systems, as they are needed for high power applications. However, there is an upper limit to the 2D carrier density which is given by the maximum electric field. This constraint holds for all known two-dimensional electron layers, such as those formed at the semiconductor-vacuum, semiconductor-oxide, and semiconductor heterostructure interfaces, and makes it rather difficult to realize very high carrier density two-dimensional electron systems, even with the application of external electric fields using gates placed very close to the surface. Despite three decades of effort, the record carrier concentrations are only in the mid-$10^{14}$ cm$^{-2}$ range.

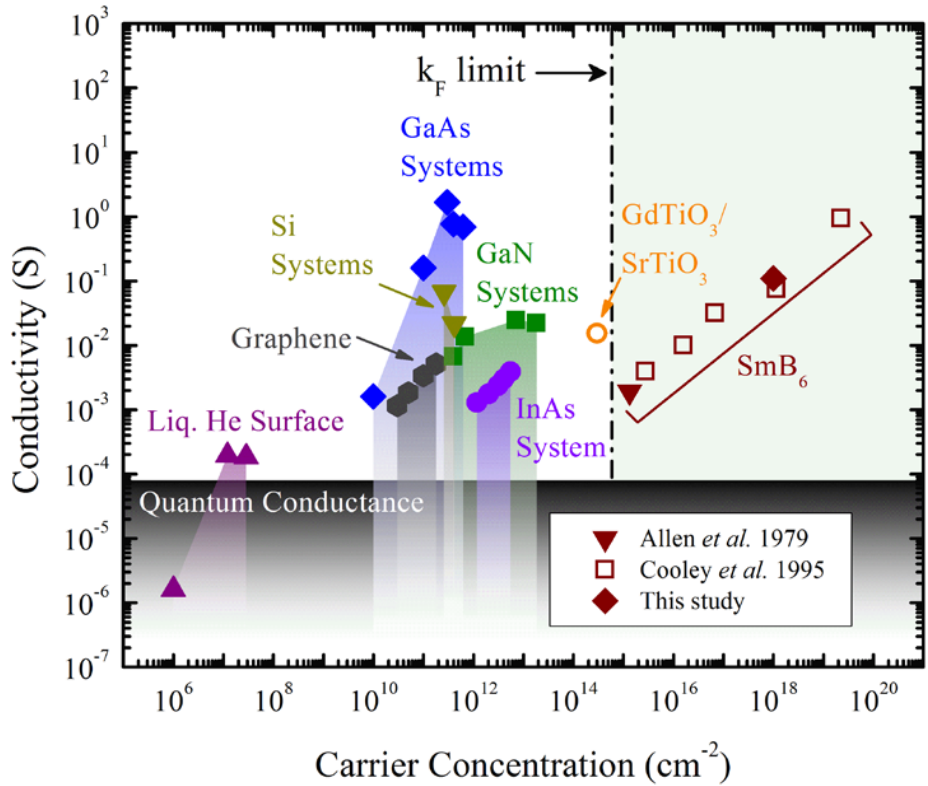

**Figure S3** – Record conductivities of electrons on the surface of Helium, GaAs/AlGaAs heterostructures, graphene, Si structures, InAs structures, GaN/AlGaN heterostructures, and a GdTiO$_3$/SrTiO$_3$ interface, plotted versus carrier concentration. The conductivity of SmB$_6$ is plotted with a single-band carrier density estimated from Hall measurements. The quantum conductance is provided as a black edge for reference. The dotted line denotes the upper $k_F$ limit imposed by the Brilliouin zone.

It is instructive to compare the surface conduction of SmB$_6$ with that of semiconductor heterostructures and other two-dimensional electron systems. In Fig. S3, we reinterpret the low T data in the SmB$_6$ literature [S1, S2] as a single-band 2D inversion layer, and compare it with that of electrons on the surface of liquid He [S3,S4], GaAs/AlGaAs heterostructures [S5 – S7], graphene [S8], Si-based structures [S9,S10], low-bandgap semiconductors such as InAs [S11], GaN/AlGaN heterostructures [S12 – S14], and a GdTiO$_3$/SrTiO$_3$ interface [S15]. The conductivities achieved in most of these interfaces, particularly the III-V semiconductor heterostructures, can only be achieved by growing the highly-engineered samples in the ultra-clean environments of molecular beam epitaxy systems, and the electron gas is typically buried well below the surface for protection. In contrast, in all existing literature on SmB$_6$, the residual conductivity persists, regardless of the quality of the sample or its surface, or of the method of preparation.

This plot conveys two general messages. First, the single-band carrier densities estimated here (particularly those under high pressure) exceed the record carrier densities of conventional inversion layers limited by the electric field to the $10^{13}$ - $10^{14}$ cm$^{-2}$ range due to the local charge density. Second, these carrier densities extend well above the $k_F$ limit imposed on single-band systems by the size of the Brillouin zone. Carrier densities of inversion layers with multiple bands can exceed the $k_F$ limit, but not by the orders of magnitude shown. We conclude that if the conduction is truly 2D, the carrier density cannot be extracted from the Hall coefficient in the usual way, as discussed below. This comparison also rules out the possibility that the surface conduction of SmB$_6$ is due solely to an ordinary band inversion or similar polarity-driven mechanisms.

It is worthwhile to consider the formation of an ordinary 3D conducting layer after polishing and etching of the surface (as we have also done in the present study). In such a scenario the conducting layer needs to be thick in order to explain the high surface conductivity in this material. The Hall measurements can be used to determine the single-band sheet carrier density $n_{3D}t$, where $n_{3D}$ and $t$ are the carrier concentration and the thickness of the surface-conducting layer, respectively. Pressure studies of the Hall conductivity [S2] for T above that of the saturation show changes consistent with the gap gradually closing, but for the low T saturation show orders-of-magnitude changes; this would require orders-of-magnitude changes in $t$, e.g., from a few nanometers to a micron. Of course, such a variation in the thickness is totally unphysical, and we can safely rule out conduction by a 3D metallic layer such as aluminium.

The possibility of ambipolar conduction indicated by topological insulator theory provides a compelling explanation of the Hall effect data we and others have observed. In this case, the Hall coefficient can be suppressed or undergo a sign change, depending on the detailed balance of electrons and holes (and their respective mobilities) in the system. Thus, the changes in the Hall coefficient as pressure closes the gap may indicate changes in this detailed balance. Such ambipolar conduction is not supported by inversion layers, polarity-driven states, or 3D metal layers.

For materials with a robust surface state, an additional suppression of the Hall coefficient may arise from the geometry of samples typically used in the Hall measurement. Although the top and bottom surfaces are perpendicular to the magnetic field, there are also side surfaces parallel to the magnetic field. The top and bottom surfaces can be considered as two parallel Hall bars, but the side surfaces, which show no Hall effect, add a metallic shell to the edges of these two Hall bars and suppress the Hall voltage significantly. Indeed, there is no possible Hall bar geometry for topological surface states. With these considerations in mind, the values obtained from past and present Hall measurements only provide us with an upper bounds for the real carrier densities. Thus, the carrier densities of $SmB_6$ plotted in Fig. S3 are only useful for comparison to ordinary conduction mechanisms, not in the context of topological insulators. Other measurements, such as Shubnikov-de Haas oscillations, must be taken to determine the carrier densities of individual bands that comprise the system.

SUPPLEMENTARY REFERENCES